\documentclass{article}  
\usepackage{bigsky2009}
\usepackage{graphicx}
\frompage{000} \topage{000}                                              
\def\rightmark{Results from Au+Au collisions at
$\sqrt{s_{NN}}$ = 9.2 GeV}
\def\leftmark{D. Das et al.(STAR Collaboration)}
 
\title{Recent results from STAR experiment in Au+Au collisions at
$\sqrt{s_{NN}}$ = 9.2 GeV} 
\authors{
{Debasish Das (for the STAR Collaboration)$^1$  %
}\\[2.812mm]
{\normalsize
\hspace*{-8pt}$^1$ Physics Department, University of California, Davis, CA 95616, USA.\\[0.2ex] 
Email : debasish@rcf.rhic.bnl.gov , debdas@ucdavis.edu \\
}}
 \abstract{Theoretical models suggest that the Quantum Chromo-Dynamics (QCD) 
phase diagram has a critical point demarcating the order of transition between the two phases: 
the hadron gas, in which the quarks are confined and the
Quark-Gluon Plasma (QGP). The central goal of the experiments with 
relativistic heavy-ion collisions is to create and study such form of 
matter called the QGP and understand the QCD phase diagram. The STAR (Solenoidal Tracker At RHIC) 
detector is pertinent for the RHIC (Relativistic Heavy Ion Collider) energy scan program where we plan to 
explore this exciting physics possibility using heavy-ion collisions 
at various center of mass energies. A first test run with Au+Au collisions at 
$\sqrt{s_{NN}}$ = 9.2 GeV took place in early 2008. We present the 
recent STAR results from this run of the identified particles (pion,
kaon and proton) transverse momentum spectra and ratios. Also we shall
present and discuss the results of the azimuthal anisotropy parameters ($v_{1}$, $v_{2}$) along
with the pion interferometry measurements. These recent results from Au+Au
collisions at $\sqrt{s_{NN}}$ = 9.2 GeV are compared with other SPS
and RHIC measurements.
}

\keyword{QGP, RHIC, QCD Phase Diagram, Critical Point} 
\PACS{25.75.N -q}

\begin{document}
 
\maketitle
\setcounter{page}{1}
\section{Introduction: }

Developing an understanding of the deconfining phase transition in hadronic matter~\cite{Wilczek:1999ym} and
of the properties of QGP (which is regarded as thermodynamic equilibrium system) has proven to be a challenging task. Lattice QCD~\cite{Karsch:2001cy} 
predicts a phase transformation to a quark-gluon plasma at a temperature of approximately 
$T \approx 170 MeV (1~MeV \approx 1.1604 \times 10^{10}K)$ 
corresponding to an energy density $\epsilon \approx 1~GeV/fm^{3}$, which is 
larger than the energy density of normal nuclear matter.

\begin{figure}[htbp]
\begin{center}
\includegraphics[scale=1.0]{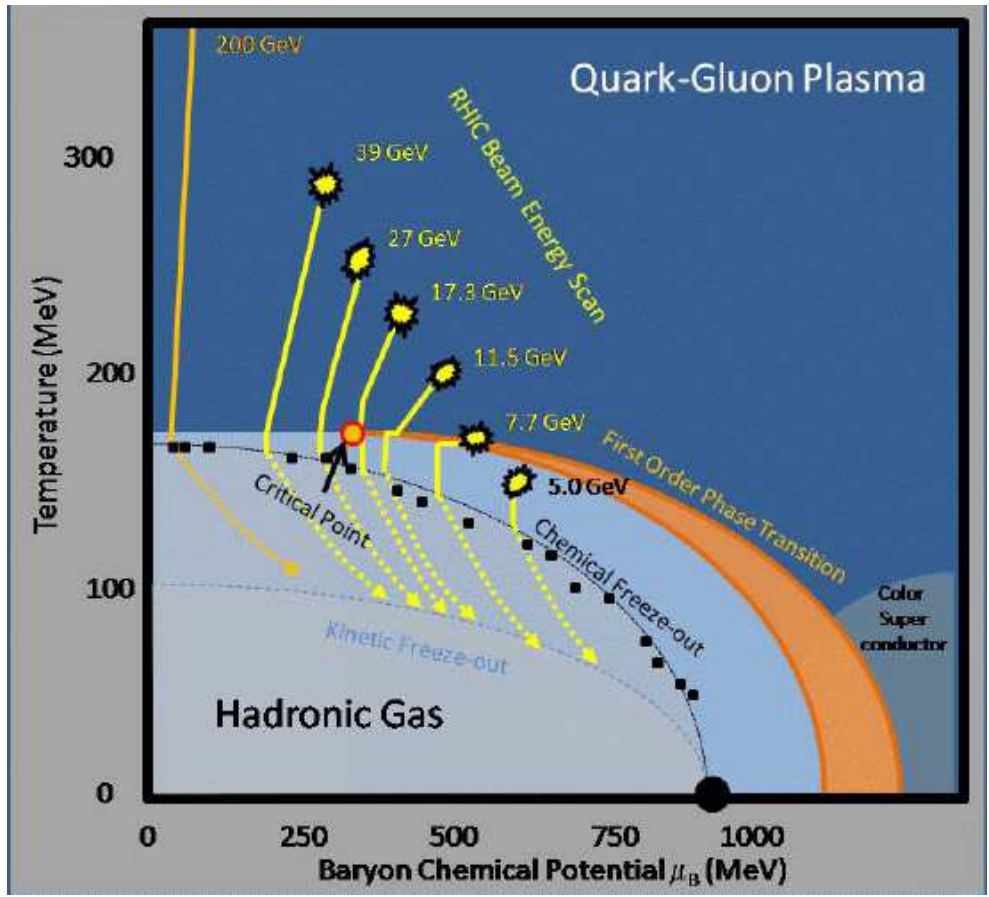}
\caption{\label{fig:phase1_ch1} (Color Online) A schematic representation of the QCD phase diagram with temperature T vs. baryonic chemical potential 
$\mu_B$. }
\end{center}
\end{figure}

The thermodynamics of quarks and gluons can be summarized as: statistical QCD predicts
the hadron-quark deconfinement transition and the properties of QGP  through the first principle
calculations of ever increasing precision~\cite{Satz:1997bs}. 
The thermodynamical information is presented in the form of a phase diagram, in which the different
manifestations and phases of a substance occupy the different regions of a plot whose axes
are calibrated in terms of the external conditions or control parameters~\cite{Hands:2001ve}. 
The phase diagram of QCD is shown in (Fig.~(\ref{fig:phase1_ch1})) where the 
the control parameters are the temperature T and the baryon chemical potential $\mu_B$. The 
freeze-out points are determined from the thermal model analyses of heavy ion collision data at SIS, AGS and 
SPS energy.

The phase diagram in Fig.~(\ref{fig:phase1_ch1}) suggests a possible occurrence of 
deconfined phases of quark matter at two extreme conditions and the chemical freeze-out values 
from Ref.~\cite{Becattini:2007ci}.
The first situation occurs when the temperature is high and net baryon density is zero.
The estimated critical temperature (at zero baryon density) is about 
170 MeV~\cite{Karsch:2001cy}. The second situation occurs when the temperature
is zero and the baryon density is about 5 times the equilibrium nuclear matter density.
The astro-physics of neutron stars provides a good testing ground for the exploration
of this very dense matter. Neutron stars are cold on the nuclear scale and have
temperatures in the range of $10^{5}$ to 
$10^{9} K$.

For a system in between these two limits, there is a pressure arising from the
thermal motion of the particles and there is also a pressure arising from the degeneracy
of the fermion gas. The total pressure is hence the sum of the two contributions.
Thus, for a system whose temperature and net baryon density is non-zero, the critical
temperature at which the quark matter becomes deconfined shall be placed between the 
limits of T = 0 for a degenerate quark gas, and the other limit of critical temperature
($T_{C}$) for a pure plasma with no net baryon density.
An important objective of modern nuclear physics is to explore the QCD phase diagram 
in the various temperatures and baryon density regions so as to confirm the
existence of the new phase of quark matter.

\begin{figure}
\begin{center}
\includegraphics[scale=.65]{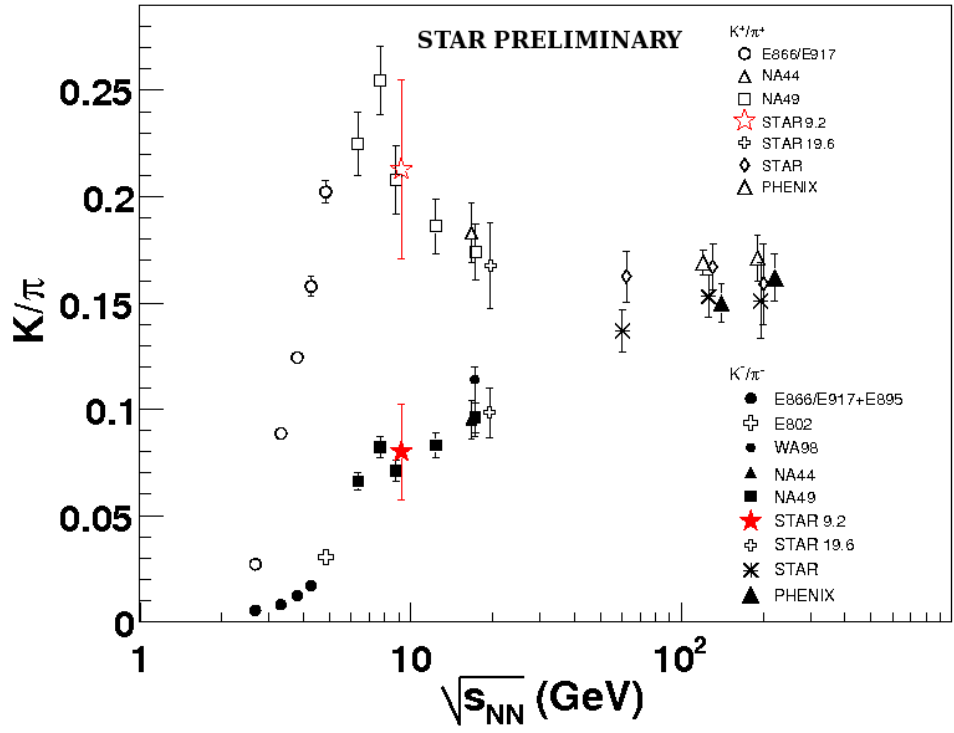}
\caption{\label{fig:kpi} (Color Online) Charged kaons to charged pions ratio as a function of
colliding beam energy. 
}
\end{center}
\end{figure}

Theoretical predictions have suggested that the QCD phase diagram has a critical point
demarcating the order of transition between the two phases: the hadron gas, in which the quarks are confined and the
QGP . The central goal of the relativistic heavy-ion collision experiments is to 
map the QCD phase diagram~\cite{Adams:2005dq} and the QCD critical
point (where the first order phase transition ends) as marked in Fig.~(\ref{fig:phase1_ch1}).
To map and understand the QCD phase diagram we need to find a way to vary temperature and $\mu_{B}$, which
can be achieved by varying the colliding beam energy. To achieve these goals, STAR has proposed a beam energy 
scan program at RHIC spanning beam energies from 5 GeV to 50 GeV. As a first step of the energy scan program, a test run
was organized in early 2008 with Au+Au collisions at $\sqrt{s_{NN}}$ = 9.2 GeV.

\section{Analysis and Results}

The data presented here are from Au+Au collisions at $\sqrt{s_{NN}}$ = 9.2 GeV using the
Time Projection Chamber (TPC) in the STAR experiment at RHIC.
The events with a primary vertex within $\pm$ 75 cm of the geometric center of TPC along
the beam axis were accepted for this analysis and about $\sim$3000 events were analyzed. 
The total systematic error for pions were found to be $\sim$10\%, $\sim$12\% for kaons and greater than 15\% for protons. 
The details of the analysis cuts and centrality selection are explained in Ref.~\cite{Kumar:2008ek}.

\subsection{Hadron Ratios}

The non-monotonic behavior of the $K^{+}\slash\pi^{+}$ as measured by NA49~\cite{:2007fe} has led to 
intense theoretical activities in recent times~\cite{Andronic:2008gu}. It has been suggested that a 
transition to a deconfined state of matter may cause anomalies in the energy dependence of pion to kaon yield ratios 
and strangeness production~\cite{:2007fe} in relativistic nucleus-nucleus collisions. The $K^{-}\slash K^{+}$ ratio 
at $\sqrt{s_{NN}}$ = 9.2 GeV and lower energies (AGS and SPS) are much less than unity and hence indicate a significant 
contribution to kaon production from associated mechanism. But for higher colliding energies (RHIC) the $K^{-}\slash K^{+}$ 
ratio is closer to unity due to the dominance of pair production mechanism for kaon production. However the $\pi^{-}\slash\pi^{+}$ 
values for $\sqrt{s_{NN}}$ = 9.2 GeV is close to unity. For lower AGS energies due to contribution from resonance decays the values 
deviate to higher numbers than unity. Figure~\ref{fig:kpi} shows the $K^{\pm}\slash\pi^{\pm}$ ratio as a function 
of $\sqrt{s_{\mathrm {NN}}}$~\cite{Kumar:2008ek}. 

\subsection{Azimuthal Anisotropy}

\begin{figure}
\begin{center}
\includegraphics[scale=.45]{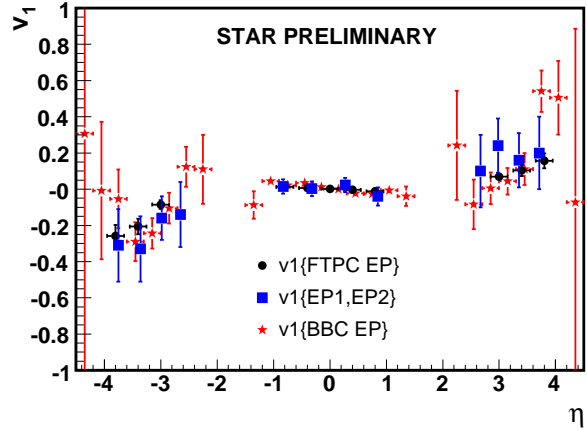}
\caption{\label{fig:v1} (Color Online) $v_{1}$ as function of $\eta$ for charged hadrons. The ``circles'' denote the use of Event Plane (EP) from FTPCs, 
``squares'' for using both FTPCs and TPC also, whereas the ``star symbol'' shows the use of BBC as EP. 
}
\end{center}
\end{figure}

The shape of $v_{1}$ with rapidity has been of interest since it can predict the nature of phase transition~\cite{Stoecker:2004qu} 
exhibiting a characteristic ``wiggle'' in proton flow. 
The preliminary results of $v_{1}$  for Au+Au collisions at $\sqrt{s_{NN}}$ = 9.2 GeV 
have been obtained using 3 different Event Planes (EP) and the results are
shown in Fig.~\ref{fig:v1} for different regions of $\eta$. 
It has been shown in Ref.~\cite{Kumar:2008ek} that the rapidity dependent behavior of $v_{1}$ 
for 9.2 GeV collisions is different from the corresponding results at 200
and 62.4 GeV. This could be due to the spectators effect, since the beam rapidity for 
9.2 GeV is 2.3, which is within the Forward Time Projection Chamber (FTPC) acceptance in the STAR experiment, whereas 
the beam rapidities are 5.4 and 4.2 for 200 GeV and 62.4 GeV respectively.

\begin{figure}
\begin{center}
\includegraphics[scale=0.45]{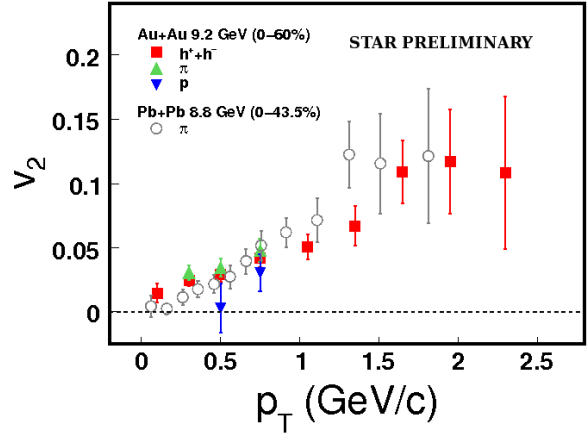}
\vspace{-0.5cm}
\caption{\label{fig:v2} (Color Online) $v_{2}$ as a function of $p_T$ for charged
hadrons (squares), pions(upper-triangles) and 
protons (lower-triangles) in 0-60$\%$ Au+Au collisions at $\sqrt{s_{NN}}$ = 9.2 GeV.
The error bars are shown only for the statistical uncertainties. For
comparison, $v_{2}$($p_{T}$) results for pions (circles) from
NA49~\cite{v2NA49} in 0-43.5$\%$ Pb + Pb collisions at $\sqrt{s_{NN}}$ =
8.8 GeV are shown.}
\end{center}
\end{figure}

Figure~\ref{fig:v2} shows $v_{2}$ ($p_{T}$ ) for charged hadrons, pions and
protons for Au+Au collisions $\sqrt{s_{NN}}$ = 9.2 GeV. 
Results are compared with pion $v_{2}$  results from the NA49 experiment at 
similar energy and are seen to compare well~\cite{Kumar:2008ek}.

\subsection{Pion Interferometry}

\begin{figure}
\begin{center}
\includegraphics[scale=0.3]{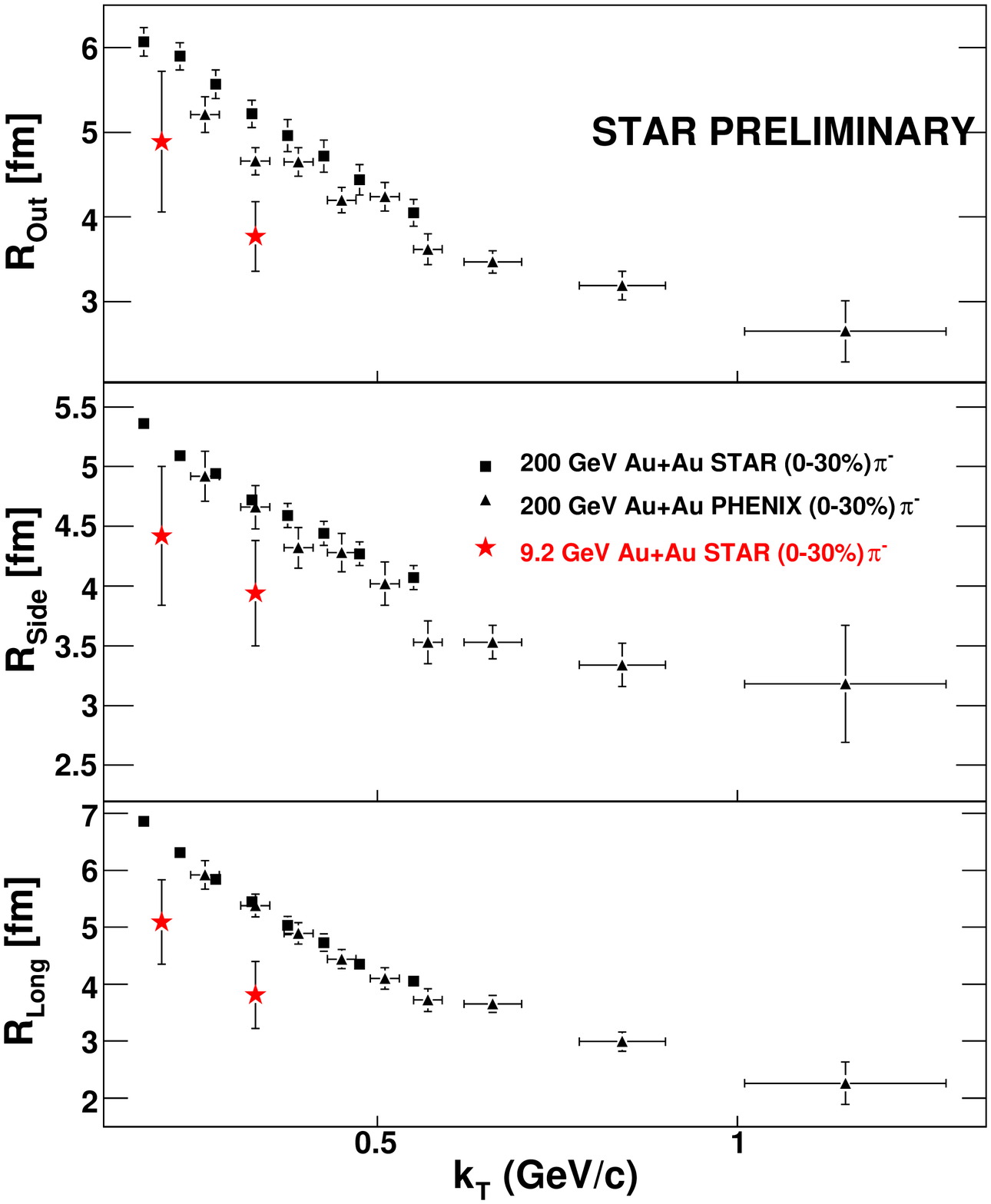}
\vspace{-0.4cm}
\caption{\label{fig:hbt} (Color Online) The HBT radii $R_{out}$, $R_{side}$ and $R_{long}$ vs. $k_{T}$.
}
\end{center}
\end{figure}

The negative pion interferometry results for Au+Au collisions at $\sqrt{s_{NN}}$ = 9.2 GeV are presented in Fig.~\ref{fig:hbt} 
and compared with 200 GeV~\cite{Adams:2004yc} measurements. 
The ratio $R_{out}/R_{side}$ is close to 1 as shown in Ref.~\cite{Kumar:2008ek}.

\section{Conclusions and Outlook}

The hadronic yields and ratios measured for Au+Au at $\sqrt{s_{NN}}$ = 9.2 GeV data are consistent with the observed beam 
energy dependence within errors~\cite{Kumar:2008ek}.
The coefficients of azimuthal anisotropy, $v_{1}$ and $v_{2}$, for Au+Au collisions
at $\sqrt{s_{NN}}$ = 9.2 GeV are presented. The $v_{2}$ results compare well with the
results from SPS experiments and follow the observed dependence on $\sqrt{s_{NN}}$. It is observed that results for
pion interferometry measurements for Au+Au 9.2 GeV also follow the established beam energy
dependence and  $R_{out}/R_{side}$ $\sim$ 1~\cite{Kumar:2008ek}. The decrease of HBT radii with $k_{T}$ is consistent with 
collective flow~\cite{Adams:2004yc}.

To conclude we have obtained and presented the results for identified hadron spectra, azimuthal
anisotropy and pion interferometry measurements for Au+Au collisions 
at $\sqrt{s_{NN}}$ = 9.2 GeV.
The few thousand events recorded during this short beam development test run in 2008, provide 
the necessary evidence that STAR detector is fully ready and capable of successful operation at sub-injection energies. 
The yellow lines in Fig.~(\ref{fig:phase1_ch1}) represent the reaction trajectories at energies($\sqrt{s_{NN}}$ = 5, 7.7, 11.5, 17.3, 27 and 39 GeV) 
proposed for the first Beam Energy Scan(BES) program.
The higher statistics and good particle identification capability in the STAR experiment 
in a Collider set-up will help in locating the critical point and  map the QCD phase diagram along with
some new measurements that were not possible in SPS. 
The lower $\sqrt{s_{NN}}$ energies shall provide the potential for the exciting critical point search whereas the higher energies 
shall bridge the current gap between RHIC and SPS energies. This will surely allow us to explore the pertinent queries about the 
``turn on/off'' effects of the signatures of partonic media such as constituent quark scaling of $v_{2}$~\cite{qscale}, 
hadron suppression~\cite{suppression} and ridge formation~\cite{ridge}.


\end{document}